# Constraint relations between the unknown coefficients in the scaled boundary finite element formulation in Electromagnetics

V.S.Prasanna Rajan*, K.C.James Raju

School of Physics, University of Hyderabad, Hyderabad - 500 046, India

**Abstract** : The constraint equations for the novel scaled boundary finite element method in electromagnetics to prevent the spurious modes in the eigen solution, is derived from the Maxwell's divergence equation for the magnetic field and by combining it with the scaled boundary transformation of the geometry.

**Introduction** :  The scaled boundary finite element method is a novel finite element method, and its theory has been recently extended to electromagnetics [1]. This scaled boundary finite element approach , was initially developed by Chongmin Song and John.P.Wolf [2-14] to successfully solve elastodynamic and allied problems in civil engineering.   The advantages of this novel finite element method has already been described in detail [1-14].

One of the crucial aspects to be taken care during the application of the finite element methods for analysis is the occurrence of the spurious modes in the finite element eigen value equation. In electromagnetics, it has been found that the occurrence of these modes is due to the inadequate modeling of the zero divergence of the magnetic field [15] . Among the various approaches suggested to avoid these modes in computation, it has been established that the use of the tangentially continuous vector finite element method greatly reduces the occurrence of these modes [15] .

___________________________________________

Corresponding author: vsprajan@yahoo.com, kcjrsprs@uohyd.ernet.in

In this situation, it is imperative to derive the conditions for the scaled boundary finite element method, so that spurious modes do not occur in the solution.

**Theory : Derivation of the constraint equations :**

The Maxwell's divergence equation for the magnetic field is given by,

Div **B** = 0      (1)

For the case where $\mu_r=1$ the above equation can be written as,

Div **H** = 0      (2)

Rewriting the above equation using the scaled boundary transformation [1],

$$\frac{g^\xi}{|J|}n_x^\xi \frac{\partial H_\xi}{\partial \xi} + \frac{1}{\xi}\left(\frac{g^\eta}{|J|}n_x^\eta \frac{\partial H_\xi}{\partial \eta} + \frac{g^\zeta}{|J|}n_x^\zeta \frac{\partial H_\xi}{\partial \zeta}\right) + \frac{g^\xi}{|J|}n_y^\xi \frac{\partial H_\eta}{\partial \xi} + \frac{1}{\xi}\left(\frac{g^\eta}{|J|}n_y^\eta \frac{\partial H_\eta}{\partial \eta} + \frac{g^\zeta}{|J|}n_y^\zeta \frac{\partial H_\eta}{\partial \zeta}\right) +$$

$$\frac{g^\xi}{|J|}n_z^\xi \frac{\partial H_\zeta}{\partial \xi} + \frac{1}{\xi}\left(\frac{g^\eta}{|J|}n_z^\eta \frac{\partial H_\zeta}{\partial \eta} + \frac{g^\zeta}{|J|}n_z^\zeta \frac{\partial H_\zeta}{\partial \zeta}\right) = 0 \quad \ldots(3)$$

Multiplying both sides of (3) by $\xi$,

$$\xi\frac{g^\xi}{|J|}n_x^\xi \frac{\partial H_\xi}{\partial \xi} + \left(\frac{g^\eta}{|J|}n_x^\eta \frac{\partial H_\xi}{\partial \eta} + \frac{g^\zeta}{|J|}n_x^\zeta \frac{\partial H_\xi}{\partial \zeta}\right) + \xi\frac{g^\xi}{|J|}n_y^\xi \frac{\partial H_\eta}{\partial \xi} + \left(\frac{g^\eta}{|J|}n_y^\eta \frac{\partial H_\eta}{\partial \eta} + \frac{g^\zeta}{|J|}n_y^\zeta \frac{\partial H_\eta}{\partial \zeta}\right) +$$

$$\xi\frac{g^\xi}{|J|}n_z^\xi \frac{\partial H_\zeta}{\partial \xi} + \left(\frac{g^\eta}{|J|}n_z^\eta \frac{\partial H_\zeta}{\partial \eta} + \frac{g^\zeta}{|J|}n_z^\zeta \frac{\partial H_\zeta}{\partial \zeta}\right) = 0 \quad \ldots(4)$$

Using the expansions for $H_\xi$, $H_\eta$, $H_\zeta$ [1]

$$H_\xi = f_1(\xi)\sum_{i=0}^{m}\sum_{j=0}^{n} h_{\xi(i,j)} h_{\xi i}(\eta) h_{\xi j}(\zeta) \qquad (5a)$$

$$H_\eta = f_2(\xi)\sum_{i=0}^{m}\sum_{j=0}^{n} h_{\eta(i,j)} h_{\eta i}(\eta) h_{\eta j}(\zeta) \qquad (5b)$$

$$H_\zeta = f_3(\xi)\sum_{i=0}^{m}\sum_{j=0}^{n} h_{\zeta(i,j)} h_{\zeta i}(\eta) h_{\zeta j}(\zeta) \qquad (5c)$$

where the functions $f_1$ $f_2$ and $f_3$ are unknown radial functions depending on the radial coordinate $\xi$, $h_{\xi(i,j)} h_{\eta(i,j)} h_{\zeta(i,j)}$ are unknown coefficients, and $h_i(\eta), h_j(\zeta)$ are the single variable functions of $\eta, \zeta$ representing the variations in ç and ærespectively and m < n and m ≠ 0 and n ≠ 0

Substituting (5) in (4) and grouping the terms involving $\xi$ and integrating both sides with respect to the two circumferential coordinates $\eta$ and $\zeta$,

$$\left(\xi\frac{df_1}{d\xi}k_1\right)+(k_2+k_3)f_1+\left(\xi\frac{df_2}{d\xi}k_4\right)+(k_5+k_6)f_2+\left(\xi\frac{df_3}{d\xi}k_7\right)+(k_8+k_9)f_3=0 \quad ....(6)$$

The expressions for $k_1$ to $k_9$ are given as follows :

$$k_1 = \sum_{i=0}^{m}\sum_{j=0}^{n} h_{\xi(i,j)} k_{1(i,j)} \qquad ...(7a)$$

$$k_2 + k_3 = \sum_{i=0}^{m}\sum_{j=0}^{n} h_{\xi(i,j)} (k_2+k_3)_{(i,j)} \qquad ...(7b)$$

$$k_4 = \sum_{i=0}^{m}\sum_{j=0}^{n} h_{\eta(i,j)} k_{4(i,j)} \qquad ...(7c)$$

$$k_5 + k_6 = \sum_{i=0}^{m}\sum_{j=0}^{n} h_{\eta(i,j)} (k_5+k_6)_{(i,j)} \qquad ...(7d)$$

$$k_7 = \sum_{i=0}^{m}\sum_{j=0}^{n} h_{\zeta(i,j)} k_{7(i,j)} \qquad ...(7e)$$

$$k_8 + k_9 = \sum_{i=0}^{m}\sum_{j=0}^{n} h_{\zeta(i,j)} (k_8+k_9)_{(i,j)} \qquad ...(7f)$$

The expressions for $(k_n)_{(i,j)}$ n= 1 to 9 in (7) are given as follows:

$$k_{1(i,j)} = \int_{(\eta_1,\zeta_1)}^{(\eta_2,\zeta_2)} \frac{g^\xi}{|J|} h_{\xi i}(\eta) h_{\xi j}(\zeta) d\eta d\zeta \qquad (8a)$$

$$(k_2^- + k_3)_{i,j} = \int_{(\eta_1,\zeta_1)}^{(\eta_2,\zeta_2)} \frac{1}{|J|}[(g^\eta n_x^\eta h'_{\xi i}(\eta) h_{\xi j}(\zeta)) + (g^\zeta n_x^\zeta h_{\xi i}(\eta) h'_{\xi j}(\zeta))] d\eta d\zeta \qquad (8b)$$

$$(k_4)_{i,j} = \int_{(\eta_1,\zeta_1)}^{(\eta_2,\zeta_2)} \frac{1}{|J|} (g^\xi n_y^\xi h_{\eta i}(\eta) h_{\eta j}(\zeta)) d\eta d\zeta \qquad (8c)$$

$$(k_5 + k_6)_{i,j} = \int_{(\eta_1,\zeta_1)}^{(\eta_2,\zeta_2)} \frac{1}{|J|}[(g^\eta n_y^\eta h'_{\eta i}(\eta) h_{\eta j}(\zeta)) + (g^\zeta n_y^\zeta h_{\eta i}(\eta) h'_{\eta j}(\zeta))] d\eta d\zeta \qquad (8d)$$

$$(k_7)_{i,j} = \int_{(\eta_1,\varsigma_1)}^{(\eta_2,\varsigma_2)} \frac{1}{|J|} (g^\xi n_z^\xi h_{\varsigma i}(\eta) h_{\varsigma j}(\varsigma)) d\eta d\varsigma \tag{8e}$$

$$(k_8 + k_9)_{i,j} = \int_{(\eta_1,\varsigma_1)}^{(\eta_2,\varsigma_2)} \frac{1}{|J|} [(g^\eta n_z^\eta h'_{\varsigma i}(\eta) h_{\varsigma j}(\varsigma)) + (g^\varsigma n_z^\varsigma h_{\varsigma i}(\eta) h'_{\varsigma j}(\varsigma))] d\eta d\varsigma \tag{8f}$$

In all the above expressions the upper and lower limits of $\eta$ and $\varsigma$ refers to the limits of $\eta$ and $\varsigma$ for a single surface element when the surface integration is performed for every element. Also, in all the above expressions(8a-8f), h' denotes the derivative of h with respect to the variable in the curved bracket. The subscripts denote the respective component terms of $\mathbf{H}(\xi,\eta,\varsigma)$.

Expanding the radial functions $f_1(\xi)$, $f_2(\xi)$, $f_3(\xi)$ as a power series in $\xi$ as

$$f_1(\xi) = \sum_{k=0}^{N} a_k \xi^k \tag{9a}$$

$$f_2(\xi) = \sum_{k=0}^{N} b_k \xi^k \tag{9b}$$

$$f_3(\xi) = \sum_{k=0}^{N} c_k \xi^k \tag{9c}$$

Substituting (7) and (9) in (6), regrouping like terms of $\xi$ and enforcing the condition on the resulting expression that **it holds for arbitrary** $\xi$, the following relations can be obtained.

$$a_0(k_2 + k_3) + b_0(k_5 + k_6) + c_0(k_8 + k_9) = 0 \tag{10a}$$

For $k > 0$,
$$a_k[(k.k_1) + (k_2 + k_3)] + b_k[(k.k_4) + (k_5 + k_6)] + c_k[(k.k_7) + (k_8 + k_9)] = 0 \tag{10b}$$

Since $a_k$, $b_k$, $c_k$ are arbitrary, they can be replaced by $h_{\xi\,(i,j)}$, $h_{\eta(i,j)}$, $h_{\varsigma(i,j)}$ respectively in the same sequence as the $h_{(i,j)}$ coefficients appear in the double summation series. The number of unknown coefficients in the double summation series for every single component of **H** is chosen to be equal to the number of unknown coefficients in the corresponding radial expansion. This results in the expression of N in terms of m and n given above. The effect of this replacement makes the radial expansion also in terms of the unknown $h_{(i,j)}$ coefficients. Following this procedure in (10) and using (8), the following constraint relations are obtained.

$$\sum_{i=0}^{m}\sum_{j=0}^{n}(h_{\xi(i,j)}h_{\xi(0,0)})[(k_2+k_3)_{i,j}] + (h_{\eta(i,j)}h_{\eta(0,0)})[(k_5+k_6)_{i,j}] + (h_{\zeta(i,j)}h_{\zeta(0,0)})[(k_8+k_9)_{i,j}] = 0$$

for $k = 0$  ....(11a)

$$\sum_{i=0}^{m}\sum_{j=0}^{n}(h_{\xi(i,j)}h_{\xi(i_k,j_k)})[(k \cdot k_{1(i,j)}) + (k_2+k_3)_{i,j}] + (h_{\eta(i,j)}h_{\eta(i_k,j_k)})[(k \cdot k_{4(i,j)}) + (k_5+k_6)_{i,j}]$$
$$+ (h_{\zeta(i,j)}h_{\zeta(i_k,j_k)})[(k \cdot k_{7(i,j)}) + (k_8+k_9)_{i,j}] = 0 \qquad \text{for } k > 0 \text{ ....(11b)}$$

where $h_{\xi(0,0)}$, $h_{\eta(0,0)}$ and $h_{\zeta(0,0)}$ correspond to the unknown h coefficients for i=j=0 for $H_\xi$, $H_\eta$, $H_\zeta$ respectively.

$h_{\xi(i_k,j_k)}, h_{\eta(i_k,j_k)}, h_{\zeta(i_k,j_k)}$ correspond to the unknown h - coefficients with corresponding (i, j) values for given $k > 0$.

It is important to note that the divergence condition which in (1) and (2) is a differential relation, is now converted into a set of constraint equations on the unknown coefficients in (11) suitable for numerical implementation. The constraint equations (11) are to be necessarily applied in order to prevent the occurrence of the spurious modes.

**Conclusion :** The necessary constraint relations for the scaled boundary finite element method for preventing the occurrence of spurious modes is derived, which is suitable for numerical implementation.

**Acknowledgement :** The first author thanks Dr.John.P.Wolf of the Department of Civil Engg, Institute of Hydraulics and Energy, Swiss Federal Institute of Technology Lausanne,Switzerland for his crucial help in sending his research papers on the scaled boundary finite element method and for his helpful suggestions. The author also thanks the Council for scientific and Industrial Research (CSIR), New Delhi , India for providing the financial assistance in the form of Senior Research Fellowship in the research project sponsored by it.**References :**

1) V.S.Prasanna Rajan, K.C.James Raju, "Theoretical aspects of a Novel Scaled Boundary Finite Element formulation in Computational Electromagnetics", submitted for review to the Applied Computational Electromagnetics society.
2) Chongmin Song and John P.Wolf, "The Scaled boundary finite-element method- alias Consistent infinitesimal finite-element cell method – for elastodynamics", *Computer Methods in applied mechanics and engineering*, No.147 , pp. 329-355, 1997.
3) Chongmin Song and John P.Wolf, "Consistent Infinitesimal Finite-Element Cell Method: Three-Dimensional Vector Wave Equation", *International Journal for Numerical Methods in Engg*, Vol.39, pp.2189-2208, 1996.
4) Chongmin Song and John P.Wolf, "Consistent Infinitesimal Finite Element Cell method for incompressible medium", *Communications in Numerical Methods in Engineering,*Vol.13, pp.21-32, 1997.
5) Chongmin Song and John P.Wolf, "Unit-impulse response of unbounded medium by scaled boundary finite-element method",*Comput.Methods Appl.Mech.Engg*, 159, pp.355-367, 1998.
6) Chongmin Song and John P.Wolf, "The scaled boundary finite-element method: analytical solution in frequency domain", *Comput.Methods Appl.Mech.Engg*, 164, pp.249-264, 1998.
7) Chongmin Song and John P.Wolf, "Body loads in scaled boundary finite-element method", *Comput.Methods Appl.Mech.Engg*, 180, pp.117-135, 1999.
8) Chongmin Song and John P.Wolf,"The scaled boundary finite element method-alias Consistent infinitesimal finite element cell method –for diffusion",*International Journal for Numerical Methods in Engineering*,45, pp.1403-1431, 1999.
9) John.P.Wolf and Chongmin Song, "The scaled boundary finite element method – a primer : derivations", *Computers and Structures* , 78, pp.191-210, 2000.
10) Chongmin Song and John P.Wolf,"The scaled boundary finite-element method – a rimer: solution procedures", *Computers and Structures* , 78, pp.211-225, 2000.
11) J.P.Wolf and F.G.Huot, "On modelling unbounded saturated poro-elastic soil with the caled boundary finite element method", *Proc.of the First Asian-Pacific Congress on Computational Mechanics*, Vol.2, pp.1047-1056, November 2001.
12) Chongmin Song and John P.Wolf,"Semi-analytical representation of stress singularities as occurring in cracks in anisotropic multi-materials with the scaled boundary finite element method", *Computers and Structures* , 80,pp.183-197, 2002.
13) Andrew J. Deeks and John.P.Wolf," Stress recovery and error estimation for the scaled boundary finite element method", *Int.J.Numer.Meth.Engng*, 54, pp.557-583, 2002.